\documentclass[12pt]{article}

\pdfoutput=1

\usepackage{amssymb,graphicx}
\usepackage[intlimits]{amsmath}

\makeatletter \@addtoreset{equation}{section} \makeatother

\begin{document}

\thispagestyle{empty}

\begin{flushright}\footnotesize

\texttt{ICCUB-16-014}
\vspace{0.6cm}
\end{flushright}

\renewcommand{\thefootnote}{\fnsymbol{footnote}}
\setcounter{footnote}{0}

\def\caln{\mathcal{N}}
\def\cL{\mathcal{L}}

\begin{center}
{\Large\textbf{\mathversion{bold}  D branes in background  fluxes \\ and Nielsen-Olesen instabilities
}
\par}

\vspace{0.8cm}

\textrm{Jorge~G.~Russo}
\vspace{4mm}

\textit{ Instituci\'o Catalana de Recerca i Estudis Avan\c cats (ICREA), \\
Pg. Lluis Companys, 23, 08010 Barcelona, Spain.}\\
\textit{Department de Fisica Cuantica i Astrofisica \\
and Institut de Ci\`encies del Cosmos,  \\
Universitat de Barcelona, Mart\'\i \ Franqu\`es, 1, 08028 Barcelona, Spain.}
\vspace{0.2cm}
\texttt{jorge.russo@icrea.cat}

\vspace{3mm}


\par\vspace{0.4cm}

\textbf{Abstract} \vspace{3mm}

\begin{minipage}{13cm}

In quantum field theory, charged particles with spin $\geq 1$ may become tachyonic in the present of magnetic fluxes above some critical field, signaling an instability of the vacuum. The phenomenon is generic, in particular, similar instabilities are known to exist in open and closed string theory, where a  spinning string state can become tachyonic above a critical field.
In compactifications involving RR fluxes $F_{p+2}$, the quantum states which could become tachyonic by the same Nielsen-Olesen mechanism are D$p$ branes.
By constructing an appropriate background with RR magnetic flux that takes into account back-reaction, we identify the possible tachyonic D$p$ brane states and compute the formula for the energy spectrum in a sector.  
More generally, we argue that in any  background RR magnetic flux, there are high spin D$p$ quantum states which become very light at critical fields.

\end{minipage}

\end{center}

\vspace{0.5cm}


\newpage
\setcounter{page}{1}
\renewcommand{\thefootnote}{\arabic{footnote}}
\setcounter{footnote}{0}





\def\Xint#1{\mathchoice
   {\XXint\displaystyle\textstyle{#1}}%
   {\XXint\textstyle\scriptstyle{#1}}%
   {\XXint\scriptstyle\scriptscriptstyle{#1}}%
   {\XXint\scriptscriptstyle\scriptscriptstyle{#1}}%
   \!\int}
\def\XXint#1#2#3{{\setbox0=\hbox{$#1{#2#3}{\int}$}
     \vcenter{\hbox{$#2#3$}}\kern-.52\wd0}}
\def\ddashint{\Xint=}
\def\dashint{\Xint-}

\newcommand{\bea}{\begin{eqnarray}} \newcommand{\eea}{\end{eqnarray}}
\newcommand{\n}{\nonumber}
\newcommand{\be}{\begin{equation}}
\newcommand{\ee}{\end{equation}}
\newcommand{\ba}{\begin{eqnarray}}
\newcommand{\ea}{\end{eqnarray}}

\def\sech{ {\rm sech}}
\def\p{\partial}
\def\pa{\partial}
\def\ov{\over }
\def\a{\alpha }
\def\g{\gamma}
\def\s{\sigma }
\def\td{\tilde }
\def\vp{\varphi}
\def\gd{\nu }
\def \ha {{1 \over 2}}

\def\stint{\strokedint}

\newcommand\cev[1]{\overleftarrow{#1}}


\section{Introduction}


In some theories, turning on magnetic fields can lead to Nielsen-Olesen instabilities \cite{Nielsen:1978rm,Ambjorn:1988tm}. They occur for charged particles with spin.
Their origin can be simply understood from
the energy formula for a charged particle in a uniform magnetic field $B$:
\be
E^2 =M^2 +q B(2\ell +1)- 2q  B S\ ,
\nonumber
\ee
where $M$ is the mass, $q$ is the charge, $S$ is the spin, we took the gyromagnetic factor $g=2$, and
$\ell =0,1,2,...$ is the Landau level.
For a scalar particle, $S=0$ and the gyromagnetic interaction contributes positively
to the energy. For a spin 1/2 fermion, the minimum energy occurs when $S=1/2$, in which case $E^2=M^2$.
However, for a spin 1 particle, such as the $W_\pm $ bosons, the component with spin parallel to the magnetic field, $S=1$,   would have
\be
E^2=M^2 - q B\ .
\nonumber
\ee
A tachyon instability appears if $qB>M^2$. This leads to condensation of $W$ bosons (which are then expected to form a triangular lattice of vortices, \cite{Ambjorn:1988gb,Ambjorn:1989bd}).
More generally, a spin $S$ particle with spin aligned to the magnetic field, becomes tachyonic if 
\be
E^2 =M^2 -q B(g |S|-1 )<0\quad {\rm or}\quad  qB> \frac{M^2}{g|S|-1}\ .
\nonumber
\ee
In string theory, there are particles with all  spins. This implies that the negative contribution coming from
the gyromagnetic interaction can be large, even for very weak magnetic fields.
For a given mass, the maximum spin scales like $M^2$, so the gyromagnetic interaction 
competes with the rest mass term. As a result, the value of the critical magnetic
field depends on the detailed form of the energy formula.
However, the general lesson is that we should expect
Nielsen-Olesen instabilities in any theory with
spinning objects which are charged under some flux that is turned on.
This applies to strings, D strings and, more generally, D$p$ branes, which
can also have arbitrarily large spin. 

Nielsen-Olesen instabilities have been studied for open strings  \cite{Ferrara:1993sq,Bachas:1995ik,Russo:2005za} and for closed strings
\cite{RTexactly,RTflux,Kiritsis:1995iu,RTmagnetic,Minwalla:2003hj} in various  contexts. In this paper we study, for the first time, this phenomenon in D branes. Different aspects of D branes in background RR magnetic  fluxes have been discussed in
\cite{Diengo}. Here we will  use  the same magnetic background.
In the presence of a  $F_{p+2}=dA_{p+1}$ background magnetic flux,  the energy of D$p$ branes  
gets reduced for some  angular momentum components.
Understanding the emergence of instabilities requires identifying potential tachyonic modes and a precise determination of the energy formula.

This paper is organized as follows. In section 2 we study tachyon instabilities in the superstring spectrum in the presence of an $H_3$ flux, in the context of an exact conformal string $\sigma $ model. 
The mass spectrum exhibits the appearance of tachyons above critical fields. We show that some of
these tachyon states  admit a semiclassical description in terms of classical string solutions
moving in the background flux. We close this section by discussing a supersymmetry-preserving field configuration and the corresponding  formula for the mass spectrum.
In section 3 we consider the dynamics of D$p$ branes moving in a background with $F_{p+2}$ flux. We identify a classical D$p$ brane solution which becomes tachyonic above a critical field.
Finally, in section 4, we give a more general discussion on instabilities  induced by generic flux configurations, in particular, for quantized fluxes on cycles. We briefly comment on the implications for the construction of metastable flux compactifications with broken supersymmetry.

\section{Fundamental string in  NS-NS 3-form flux $H_3$}

It is useful to start with strings in $H_3$ fluxes, as in this case we can determine the complete
energy spectrum.
We consider the  following string $\sigma $ model
\bea
L &=& -\p_+ t\p_- t +\p_+ \rho \p_-\rho + F(\rho) \rho^2 (\partial_+ \varphi   +b  \partial_+ y)(\partial_- \varphi   - b  \partial_- y) 
\nonumber\\
&+& \p_+ y\p_- y +\p_+ x^i \p_- x^i +\frac{1}{4}\alpha' \sqrt{g}R^{(2)}\big( \phi_0+{\textstyle \frac{1}{2}}\ln F \big)\ ,\qquad F=\frac{1}{1+b^2\rho^2}\ .
\label{uno}
\eea
with $i=1,...,6$. It represents an exact conformal field theory for type II superstrings. We use the conventions $\sigma^{\pm} =\tau\pm \sigma $ and
$\sigma \in [0,2\pi )$. $y$ is a periodic coordinate, 
$y=y+2\pi R\ $.
For simplicity, here we only quote the bosonic part. It is straightforward to incorporate the world-sheet fermions \cite{RTflux}.
The Lagrangian (\ref{uno}) describes strings propagating in the geometry
\bea
&& ds^2= -dt^2+d\rho^2 +\rho^2 F(\rho) d\varphi^2 +F(\rho) dy^2+dx_i^2\ ,
\nonumber\\
&& e^{2(\phi-\phi_0)} = 
F\ ,\qquad B_2= 
b \rho^2 F(\rho) \ d\varphi \wedge dy\ .  
\label{ddos}
\eea
This string model is obtained, by T-duality in the $y$ direction, from the Kaluza-Klein ``Melvin" string sigma model \cite{RTflux}
\bea
L &=& -\p_+ t\p_- t +\p_+ \rho \p_-\rho +  \rho^2 (\partial_+ \varphi   +\tilde b  \partial_+ y)(\partial_- \varphi   + \tilde b  \partial_- y) 
\nonumber\\
&+& \p_+ y\p_- y +\p_+ x^i \p_- x^i \ ,
\label{melvinRT}
\eea
with  constant dilaton and vanishing $B_2$. The $\sigma $-model (\ref{melvinRT}) describes a locally flat spacetime, as can be seen
by a formal redefinition of the polar angle, $\varphi'=\varphi+\tilde b y$. Because $y$ is periodic,
the solution of the model is  non-trivial.
The model (\ref{melvinRT}) can be solved 
in terms of free fields by carefully imposing the appropriate boundary conditions.

 It is convenient to introduce complex coordinates $x=\rho e^{i\varphi}$. 
 Clearly, $\rho e^{i\varphi'}$ is a free field and the general solution is given by  \cite{RTflux}
\bea
&& x=e^{-i \tilde b y(\tau,\sigma)} \left( X_+(\sigma ^+) +X_-(\sigma^- ) \right)\ ,\nonumber\\
&& y= y_+(\sigma_+) + y_-(\sigma_-)\ .
\label{Msolup2}
\eea
One can choose world-sheet coordinates such that 
\be
y_+(\sigma_+) +y_-(\sigma_-) = y_0+ w R \sigma + \alpha' p_y \tau\ , \quad 
\label{Msolup4}
\ee
where $w$ is the winding number and $p_y=n/R$, $n\in {\mathbb Z}$  is the quantized Kaluza-Klein momentum.
Demanding that $x$ is single-valued, $x(\sigma+2\pi,\tau)=x(\sigma,\tau)$, leads to
twisted boundary conditions for  free fields $X^+, \ X^-$, 
\be
X(\tau,\sigma +2\pi )= e^{2\pi i \gamma} X(\tau,\sigma)\ ,\ \ \ \gamma \equiv \tilde b w R \ .
\ee
Hence
\be
X_\pm=  e^{\pm i \gamma \sigma_\pm} \ \chi_\pm (\sigma_\pm ) ,
\ee
where $\chi_\pm (\sigma_\pm ) $ are single-valued free fields, with the Fourier expansion
\be
\chi_+ = i\sqrt{\frac{\alpha'}{2}} \sum _{n=-\infty}^\infty \tilde a_n \ e^{-in \sigma^+}\ ,\qquad
\chi_- = i\sqrt{\frac{\alpha'}{2}} \sum _{n=-\infty}^\infty  a_n \ e^{-in \sigma^-}\ .
\ee

The $\sigma $-model (\ref{uno}), while equivalent as a conformal field theory
to the T-dual model (\ref{melvinRT}), describes strings moving in curved space-time and as a result the general solution to the string equations of motion is more complicated. It can be obtained from section 5  in \cite{RTexactly}, adapting the parameters to the present case. 
 The  general solution is 
 \be
x=e^{-i b \left(y_+(\sigma_+) -y_-(\sigma_-)\right)} \left( X_+(\sigma ^+) +X_-(\sigma^- ) \right)\ ,\ 
\label{solup1}
\ee
\be
y= y_+(\sigma_+) + y_-(\sigma_-)-  b\tilde \varphi\ ,
\label{solup2}
\ee
\be
 \p_\pm \tilde\varphi =\pm \frac{i}{2} (X^* \p_\pm X - X \p_\pm X^*) \ .
\label{solup3}
\ee 
Note that
\be
y_+(\sigma_+) -y_-(\sigma_-) =  w R \tau + \alpha' p_y \sigma\ .
\label{solup4}
\ee
$X_+, \ X_-$ satisfy  the same
twisted boundary conditions  as before, now with 
$$
\gamma = b \alpha' p_y\ .
$$
Note, however, that $y$, given by (\ref{solup2}), is {\it not} a free field in this case.

\subsubsection*{Mass spectrum and tachyon instabilities}

The mass spectrum for the model (\ref{uno}) is  given by the formula
\be
\label{espiero}
\alpha' M^2 = 2 \hat N_R+2 \hat N_L + \frac{\alpha' n^2 }{R^2}+ \frac{ R^2}{\alpha'}  \big(w-b \frac{\alpha'}{R} \hat J\big)^2
-2\gamma (\hat J_R-\hat J_L)\ ,\ 
\ee
\be
\hat N_R- \hat N_L = n w\ ,\qquad \ \ \gamma =  \frac{\alpha' n b}{R}\ .
\nonumber
\ee
Here $\hat N_R,\ \hat N_L$ have the standard free string theory expression in terms of normal-ordered bi-linear products of bosonic and fermionic oscillators,
\be
\hat N_{R,L} = N_{R,L}- a\ ,\ \ \ \ a^{\rm (R)}=0\ ,\ \ \ a^{\rm (NS)}=\frac{1}{2}\ .
\nonumber
\ee
Thus $\hat N_{R,L} $ have eigenvalues $0,1,...$ (half-integer eigenvalues are projected out by GSO).
The angular momentum operators are
\be
\hat J_{L,R} = \pm \big( l_{L,R}+{\textstyle \frac{1}{2}} \big) +S_{L,R}\ ,\qquad \hat J= \hat J_L+\hat J_R = l_L-l_R+S_L+S_R\ ,
\nonumber
\ee
and $l_{L,R}=0,1,2,...$ are Landau quantum numbers and the spin components $S_{L,R}$ have the same expressions as in free string theory
and satisfy the bound $|S_{L,R}|\leq \hat N_{L,R}+1$.
 The mass formula is periodic with $\gamma $. As written, the formula  applies for $|\gamma |<1$; in other intervals one
must substitute $\gamma\to \hat \gamma =\gamma- [\gamma ]$.

We note that there are two gyromagnetic couplings in (\ref{espiero}): one involves the product of the winding charge $w$ and $\hat J_R+\hat J_L$; the other one involves the product of the Kaluza-Klein charge $n$ and  $\hat J_R-\hat J_L$. Equivalently, the Left charge $Q_L=\frac{wR}{\alpha'}+\frac{n}{R}$ couples to $J_R$, the Right charge $Q_R=\frac{wR}{\alpha'}-\frac{n}{R}$ couples to $J_L$.


\medskip

The mass spectrum for the string model (\ref{uno}) contains tachyons for a flux above some critical value, which are not projected out by GSO.
The first tachyon appears for the state
\bea
&& \hat N_R=\hat N_L=0\ ,  \quad S_R=1\ , \quad S_L=-1\ ,\quad w=0\ ,\quad  n=1\ ,
\nonumber\\
&&l_R=l_L=0\ ,\quad \hat J=0\ ,\quad \hat J_R-\hat J_L =1\ ,\qquad \gamma=b\, \frac{\alpha'}{R}.
\label{tach}
\eea
Then
\be
\label{dilatones}
 M^2 = \frac{1}{R^2} - 2 \frac{b}{R}\ ,
\ee
which is tachyonic for $b>b_{\rm cr}$, $b_{\rm cr} =1/(2R)$.  Now, taking into account the condition $\gamma<1$, the tachyon can appear only when
$R^2>\alpha'/2$.


Let us now consider states with high quantum numbers, $\hat N_{R}$ or $\hat N_L \gg 1 $.
In particular, we consider the state
\bea
&& \hat N_L= \hat N_R\ ,\quad w=0\ ,\quad p_y=n/R\ , 
\nonumber\\
&& l_R=l_L=0\ ,\quad S_R=-S_L=\hat N_R+1\ \ ,
\nonumber\\
&&  \hat J=0\ ,\quad \hat J_R-\hat J_L =2\hat N_R+1\ .
\label{jotas}
\eea
These are states with maximum left and right components of the spin, chosen in a way that the negative contribution to the energy coming from the gyromagnetic interaction involving $\hat J_R-\hat J_L$ is maximized. 
Substituting these quantum numbers into the mass formula,  we get
\bea
\alpha' M^2 &=& 4\hat N_R + \frac{\alpha' n^2 }{R^2} -2\gamma (2\hat N_R+1 )
\nonumber\\
&=& 4\hat N_R (1-\gamma)+ \frac{\alpha' n^2 }{R^2} -2\gamma 
\nonumber\\
&=&  \frac{\alpha' n^2}{R^2} +2 (1-\gamma)(\hat J_R-\hat J_L)- 2 \ ,\qquad \gamma = \frac{\alpha' n b}{R}\ .
\label{quantumE}
\eea
This is tachyonic for
\be
\gamma > \frac{4\hat N_R+\alpha' n^2/R^2}{4\hat N_R+2}\ .
\label{gata}
\ee
The condition $\gamma <1$ is satisfied with $n^2 < 2R^2/\alpha'$.
All these states become tachyonic above the critical field (\ref{gata}).

\subsubsection*{Classical string solutions associated with instabilities}

It is instructive to identify the  classical string solution corresponding to the tachyonic states
with $\hat N_R=\hat N_L\gg 1$, for which a semiclassical description should apply.
Since the state has maximum value for the spin components, this state 
is obtained by setting to zero all oscillator modes $a_n,\ \tilde a_n$ with $|n|\neq 1$, and turning on $\tilde a_{\pm 1},\ a_{\pm 1}$
such that the solution has maximum $\hat J_R-\hat J_L$, with vanishing total $\hat J=\hat J_R+\hat J_L$, i.e. the same quantum numbers as in (\ref{jotas}).
Note that non-zero Landau levels, coming from $a_0,\ \tilde a_0$, give an (orbital) contribution to the angular momentum in the antiparallel direction and they increase the energy.

The left and right angular momentum components are given by the following formulas (see (5.29) in \cite{RTexactly})
\be
J_R=-\frac{1}{2} \sum_n (n+\gamma ) a_n^* a_n\ ,\qquad J_L=-\frac{1}{2} \sum_n (n-\gamma ) \tilde a_n^* \tilde a_n\ .
\ee
The classical string with the same angular momentum components as the quantum state (\ref{jotas}) thus has
$\tilde a_1\neq 0$ and $a_{-1}\neq 0$, setting all other modes to zero. This gives
\be
J_R=\frac{1}{2}  (1-\gamma ) a_{-1}^* a_{-1}\ ,\qquad J_L=-\frac{1}{2}  (1-\gamma ) \tilde a_1^* \tilde a_1\ ,
\label{jori}
\ee
with $|a_{-1}|^2=|\tilde a_{1}|^2$.
Therefore, the solution (\ref{solup1}), (\ref{solup4}), with $w=0$, takes the form
\be 
x= e^{-i  b \alpha' p_y\sigma  } (X_++X_-)\ ,
\ee
where 
\be
X_+ = i\sqrt {\frac{\alpha'}{2}}\ e^{i\gamma \sigma_+} \ \tilde a_1 e^{-i \sigma_+} = L\ e^{-i(1-\gamma )\sigma_+}\ ,
\ee
\be
X_- =  i\sqrt {\frac{\alpha'}{2}} \ e^{-i\gamma \sigma_-} \ a_{-1} e^{i \sigma_-} =L\ e^{i(1-\gamma )\sigma_-} \ .
\ee
That is,
\be\label{pulsi}
x= 2L e^{-i\sigma } \cos [(1-\gamma)\tau]\ .
\ee
This represents a circular pulsating string. The classical description applies for $L\gg l_s$, where $\ l_s\equiv \sqrt{\alpha'}$.

Using (\ref{solup2}), (\ref{solup3}), we also find
\be\label{pulsi2}
y = \alpha' p_y \tau -b L^2 \left( 2(1-\gamma)\tau + \sin [2(1-\gamma ) \tau ]\right)\ .
\ee
In order to compute the energy, we have to solve the Virasoro constraints $T_{++}=T_{--}=0$.
This leads to the formula
\bea
\label{clasien}
\alpha' p_0^2 & = & \alpha' p_y^2 + \frac{4L^2}{\alpha'} \ \big( 1-\gamma \big)^2 
\nonumber\\
& = & \frac{ \alpha' n^2}{R^2}  + 2(1-\gamma) (J_R-J_L)\ .
\eea
This agrees with the quantum formula (\ref{quantumE}) except for the $-2$ contribution.
This is a small quantum contribution, which is not visible in the semiclassical approximation, which requires  $\hat N_R\gg 1$
and thus $4\hat N_R+2\approx 4\hat N_R$.
In the classical limit, the critical field gives $\gamma\to 1$ (see (\ref{gata})). As a result, the energy --which for zero field is large--
attains a minimum value $p_0= n/R $.
The tachyon instability is not seen classically, but we know that it exists, because we have computed the exact quantum spectrum.
Classically, the sign that there is an instability at a critical field is the fact that, as the magnetic field is increased, the energy decreases from a very large, macroscopic value 
$l_s p_0= 2\sqrt{N_R}=2 L/l_s \gg 1$ to a tiny value $O(l_s/R)$, i.e. to a value that may be overcome by a negative quantum contribution of $O(1/l_s)$.

\subsubsection*{Supersymmetric extension and mass spectrum}

The background (\ref{ddos}) breaks all supersymmetries.
A way to preserve a fraction of the supersymmetries is by introducing several magnetic fields in different  planes \cite{RTfluxbranes}.
For example, consider the  solution obtained by introducing two fluxes, one in the plane $(r_1,\varphi_1)$ and the other
in the plane $(r_2,\varphi_2)$.
The background is as follows 
\bea
&& ds^2=-dt^2+dx^2_i+dr_1^2+dr_2^2+r_1^2 d\varphi_1^2 +r_2^2 d\varphi_2^2 +\Lambda^{-1} \left( dy- (b_1 r_1^2 d\varphi_1+b_2 r_2^2 d\varphi_2)^2\right)
\nonumber\\
&& e^{-2(\phi-\phi_0)} = \Lambda \ ,\qquad B_2= \Lambda^{-1}  \ (b_1 r_1^2 d\varphi_1+b_2 r_2^2 d\varphi_2)\wedge dy
\nonumber\\
&& \Lambda=1+b_1 r_1^2+ b_2 r_2^2\ .
\label{fsusy}
\eea
The corresponding string sigma model is exactly conformal and solvable, and the mass spectrum is given by \cite{RTfluxbranes}
\bea
\alpha' M^2 &=& 2(\hat N_R +\hat N_L) + \frac{ R^2}{\alpha'} (w - b_1 \frac{\alpha'}{R} \hat J_1-b_2\frac{\alpha'}{R} \hat J_2)^2 + \alpha' \frac{n^2}{R^2}
\nonumber\\
&-& 2\alpha' b_1 \frac{n}{R} (\hat J_{1R}- \hat J_{1L}) - 2\alpha' b_2 \frac{n}{R} (\hat J_{2R}- \hat J_{2L}) \ ,
\label{susyspec}
\eea
with $\hat N_R-\hat N_L=nw$. For $ b_1\neq \pm b_2$, there are tachyonic instabilities appearing at critical fields.
However, when $b_1=\pm b_2$, there are sixteen unbroken supersymmetries.
In this case, one has $M^2\geq 0$ for all states in the spectrum, as expected from supersymmetry.
Taking $b_1=b_2\equiv b$, the mass
spectrum becomes
\be
 M^2 = \frac{2}{ \alpha'}(\hat N_R +\hat N_L) + \left( \frac{wR}{\alpha' } - b (\hat J_1+\hat J_2)\right)^2 +  \frac{n^2}{R^2}
- 2 b \frac{n}{R} (\hat J_{1R}+\hat J_{2R}- \hat J_{1L}- \hat J_{2L}  ) \ .
\label{DenergiaSusy}
\ee
An important class of states are the BPS states, investigated in \cite{Diengo}. 
The BPS states  have $\hat N_L=0$ (or  $\hat N_R=0$). Then the mass becomes a perfect square provided $\hat J_1+\hat J_2=\hat J_{1R}+\hat J_{2R}- \hat J_{1L}- \hat J_{2L}$,
which implies
\be
(S_{1L},S_{2L}) =(-1,0)\ \ {\rm or} \ \ (0,-1) \ .
\nonumber
\ee
Then
\be\label{bepese}
M^2_{\rm BPS}=\left( \frac{n}{R}+\frac{wR}{\alpha'} - b (\hat J_1+\hat J_2) \right)^2
\ .
\ee
Supersymmetry ensures that the energy formula is exact and does not receive quantum corrections.
In the zero field limit, this gives the familiar formula, with $M_{\rm BPS}$ being the sum of winding and KK charges. Since $\alpha' n b/R<1$, and $\hat J_1+\hat J_2\leq \hat N_R=nw$, the minimum mass is not zero, but $M_{\rm BPS}=n/R$. For $R\gg \sqrt{\alpha'}$, this is a tiny, microscopic energy, compared with the energy at zero field, 
$M_{\rm BPS} =wR/\alpha' +n/R$. This is a surprising feature, since the formula 
$\hat N_R=nw$ implies that, for $nw\gg 1$, this is highly excited string state.
Moreover,
the BPS state admits a semiclassical description in terms of a long, rotating string, where the same energy formula is reproduced \cite{Diengo}. This macroscopic string has a microscopic energy at a critical field.

\section{Dp branes in the $F_{p+2}$ background}

Consider the general problem of a D$p$ brane moving on a background with $F_{p+2}$ flux. Assuming that there is no other field turned on, the D$p$ brane action is given by
\be
S_{{\rm D}p}=-\tau_p \int d^{p+1}\sigma \ e^{-(\phi - \phi_0)} \sqrt{-{\rm det}(G_{ab})}+i\tau_p \int A_{p+1}\ \ ,
\label{Infeld}
\ee
with 
$$
\tau_p=\frac{1}{(2\pi)^p l_s^{p+1} g_s}\ ,\ \ \ l_s\equiv \sqrt{\alpha'}\ .
$$
Starting with (\ref{fsusy}), by S-duality we find a solution with $F_3=dA_2$ flux.
By T-duality transformations along the different $x_i$, one can construct a solution with $F_{p+2}$ flux,
\bea
 ds^2 &=& \Lambda^{\frac{1}{2}}(-dt^2 + dx_s^2+ dr_{1}^2+dr_2^2+r_1^2 d\varphi_1^2+r_2^2 d\varphi_2^2)
\nonumber\\
&+& \Lambda^{-\frac{1}{2}} \left(dy_1^2+...+dy_p^2-(b_1 r_1^2 d\varphi_1+b_2 r_2^2 d\varphi_2)^2\right)
\nonumber\\ 
A_{p+1} &=&  e^{-\phi_0} \Lambda^{-1} \ (b_1 r_1^2 d\varphi_1+b_2 r_2^2 d\varphi_2)\wedge dy_1\wedge ... \wedge dy_p
\nonumber\\
e^{2(\phi-\phi_0)} &=& \Lambda^{ \frac{3}{2} -\frac{p}{2}}\ , \qquad
\Lambda = 1+b_1 r_1^2+ b_2 r_2^2\ ,
\label{Sfsusyp}
\eea
with $s=p+5,...,9$.

Consider first the action (\ref{Infeld}) for the D string in the background (\ref{Sfsusyp}) with $p=1$.
One immediately finds that this coincides with the Nambu-Goto action for a fundamental string moving in the background  (\ref{ddos}).
As the Lagrangians are the same, this has the important consequence that the solutions are the same and that the {\it classical energies} for any D string motion will be the same as in the fundamental string case, with the appropriate modification in the string tension (this arises because, after S duality, one must rescale time variable in string frame by a factor $g_s$ to have Minkowski metric in the zero field limit).\footnote{The S-dual background (\ref{Sfsusyp}) is not expected to be an exact solution to all $\alpha'$ order. However,
possible $\alpha'$ corrections do not affect the energies of classical strings, which have lengths much greater than the string scale.}

In the supersymmetric $b_1=\pm b_2$ case, there is a D$p$ brane  analog of the BPS string state (\ref{bepese}),  studied in \cite{Diengo}.
The rotating D$p$ brane has essentially the same energy  formula (\ref{bepese}), replacing $wR/\alpha'$ by $\tau_p {\rm Vol}(T_p) \omega $, where $\omega $ is the winding number of the D$p$ brane around the torus $T_p$,
\be\label{otroB}
M_{\rm BPS}= \frac{n}{R}+|\tau_p {\rm Vol}(T_p) \omega  - b J |
\ .
\ee
In particular, the energy
of this macroscopic D$p$ brane decreases down to a tiny value $n/R$ at a critical field.
At the critical field, the gyromagnetic coupling cancels the huge D brane energy  $\tau_p {\rm Vol}(T_p) \omega $ due to tension.

For  $b_1\neq \pm b_2$, the spectrum may contain tachyons above some critical fields, which  will now be of order $1/g_s$.
Consider, for example, the non-supersymmetric solution with $b_2=0$, $b_1\equiv b$.
The classical solution representing a pulsating D$p$ brane, which is the analogue of the pulsating circular string solution (\ref{pulsi}), (\ref{pulsi2}) considered above, can be constructed
by starting with the following ansatz.
$$
t=\kappa\tau ,\quad r_1 = r(\tau)\ ,\quad \varphi =-\sigma_1\ ,\quad
y_1=y_1(\tau)\ ,\quad y_j =w_j R_j \sigma_j\ ,\quad j=2,...,p\ .
$$
The D$p$ brane Lagrangian thus takes the form
\be
\cL_{{\rm D}p}=-\frac{c_p\ r }{1+b^2 r^2}
\sqrt{(\kappa^2 -\dot r^2)(1+b^2 r^2)-\dot{y_1}^2}+ \frac{c_p \ br^2 \dot y_1}{1+b^2r^2}\ ,
\ee
with 
$$
c_p\equiv (2\pi)^p \tau_p    R_2...R_p w = \frac{R^{p-1} w}{l_s^{p+1} g_s}\ ,\quad w=w_2...w_p\ .
$$

The canonical momenta $p_y$ and $p_r$ are given by
\bea
&& p_y= \frac{\partial \cL_{{\rm D}p}}{\partial \dot y}= c_p \left(\frac{r\ \dot y}{(1+b^2 r^2)\sqrt{(\kappa^2 -\dot r^2)(1+b^2 r^2)-\dot{y}^2}}+ \frac{br^2}{1+b^2 r^2}\right)
\\
&& p_r=  \frac{\partial \cL_{{\rm D}p}}{\partial \dot r}=  c_p  \frac{r \ \dot r }{\sqrt{(\kappa^2 -\dot r^2)(1+b^2 r^2)-\dot{y}^2}}
\eea
The Hamiltonian is then given by
\be
\label{jamin}
H= p_r \dot r+p_y\dot y- \cL_{{\rm D}p} =  c_p \frac{r \ \kappa^2 }{\sqrt{(\kappa^2 -\dot r^2)(1+b^2 r^2)-\dot{y}^2}}\ .
\ee
One can express $\dot y$ and $\dot r$ in terms of $p_y$ and $p_r$,
\bea
&& \dot r=\frac{\kappa p_r}{\sqrt{p_y^2+ c_p^2\ r^2 (1-\gamma )^2}}\ ,
\\
&& \dot y=\frac{\kappa \big(p_y -c_p b r^2(1-\gamma )\big)}{\sqrt{p_y^2+ c_p^2\ r^2 (1-\gamma )^2}}\ ,
\eea
where 
\be
\gamma=\frac{b p_y}{c_p}\ .
\ee 
Substituting these  expressions into the Hamiltonian (\ref{jamin}), after some algebra one finally obtains the surprisingly simple formula
\be
H=\kappa \sqrt{p_r^2 + V(r)}\ ,
\label{jamil}
\ee
with
\be
V(r)=p_y^2+ c_p^2\ r^2 (1-\gamma )^2\ .
\label{armonico}
\ee
This Hamiltonian is dimensionless and generates translations in world-volume time $\tau$.
The spacetime Hamiltonian is 
$$
{\cal H}= \kappa^{-1}\ H =\sqrt{p_r^2 + V(r)}\ ;
$$ 
it has as usual dimension of energy  and generates translations in $t$.

The classical solution can be obtained by using energy conservation. It is similar to the circular string.
We find
\be
r (\tau)= 2L\cos[ (1-\gamma) \tau] \ ,\quad y = q \tau -b L^2 \left( 2(1-\gamma)\tau + \sin [2(1-\gamma ) \tau ]\right)\ ,
\label{anzos}
\ee
with $\gamma = b q$, and $q=p_y/c_p$.
The classical energy is then
\be
E^2=  \frac{n^2}{R^2}+ \frac{4w^2}{g_s^2}\ \frac{R^{2p-2}L^2}{l_s^{2p+2} }\  (1-\gamma)^2\ .
\label{spaEp}
\ee
Again the energy attains a minimum value at $\gamma \to 1$ where the D$p$ brane energy becomes tiny, equal to $p_y$ (rather than
proportional to the size $L$). The difference with the BPS D$p$ brane is that, in the BPS case, $E^2$   is garanteed to be greater or equal to zero by supersymmetry.
For non-supersymmetric configurations, there is nothing that prevents $E^2$ from receiving a small, quantum negative correction, which can cause tachyon instabilities above a critical magnetic field.

\subsubsection*{Quantum energy}

So far we have computed the classical energy (\ref{spaEp}).
Although  a complete quantization of the $p$-brane is not known for $p>1$,
 in the sector of the ansatz (where only a specific oscillation mode of the D$p$ brane is turned on)
we have a simple Hamiltonian and we can compute the  quantized eigenvalues.

Consider first the fundamental string, for which the exact quantum spectrum is known\footnote{Here, of course, we mean  the exact quantum spectrum in $\alpha'$; the quantum spectrum that includes string loops is not known even for flat spacetime.
The regime of validity of the present approximations is $g_s\ll 1$ (or $g_s\gg1$, in a dual description). That is, quantum field theory loops are not included.}
 and whereby we can  check the consistency of the  present approach.
In the Nambu-Goto formulation, the Hamiltonian is given by (\ref{jamil})
with $c_p\to 2\pi\tau_1=1/\alpha'$.

The eigenvalues of $H^2$ are well-known, since $H^2/2$ is the Hamiltonian
of the radial part of a 2-dimensional harmonic oscillator with frequency
$\omega=|1-\gamma|/\alpha'$. Assuming as before $\gamma<1$, we can remove absolute value bars.
Therefore
\be\label{nuestra}
\alpha' {\cal H}^2=\alpha' E^2 =\alpha' p_y^2+ 2(2k+1)\ \big(1-\gamma\big)+\delta
\ ,\qquad k=0,1,2,....
\ee
where $k$ is radial quantum number. 
We have added  a constant $\delta $ to account for possible normal ordering contributions, which cannot be computed from first principles in this sector. The reason is that
the normal ordering constant picks contributions from all oscillators of the string, including all those set to zero in this sector. 
It is easy to compute $\delta $ in the zero field limit. In this limit, supersymmetry must be restored.
A supersymmetric Hamiltonian must have zero energy in the ground state.
This requires $\delta\big|_{b=0}=-2$, so that $E=0$ for $k=0$, $p_y=0$.

Of course, the zero field limit does not determine what is $\delta $ at finite field $b$.
 $\delta $ could in general be $b$ dependent.
One can determine  $\delta $ by computing the ground state ($k=0$) energy at finite field.
In this  case, the string is rigid, it has no excitation, and it is described by the supergravity multiplet. The energy of the graviton can be computed by using the effective field theory,
i.e. solving the Laplace-type equation for the gravitational fluctuations in the magnetic background (\ref{ddos}) (for a  discussion, see \cite{RTmagnetic}).
For a state with $S_{R}=1$, $S_{L}=-1$, $p_y=1/R$, this calculation leads to the same formula (\ref{dilatones}) found before for $\hat N_R=\hat N_L=0$, that we reproduce here 
\be
\label{dilaga}
 E^2 = \frac{1}{R^2} - 2 \frac{b}{R}\ .
\ee
On the other hand, setting $k=0$ and $p_y=1/R$ in our formula (\ref{nuestra}), we obtain
\be
\alpha' E^2 =\frac{1}{R^2}- 2 \frac{b}{R} + 2+\delta\ ,
\ee
therefore  $\delta =-2$ for all $b$.
This matches (\ref{quantumE}),
\be
\alpha'  E^2 = \alpha'  p_y^2+ 2(2k+1)\ \big(1-\gamma\big)-2\ .
\ee
Note that the radial quantum number $k$  is related to $\hat J_R- \hat J_L$.
Therefore we find the correct formula for the quantum spectrum in this sector.

We now generalize this calculation for a D$p$ brane.
 The Hamiltonian (\ref{jamil}), (\ref{armonico}) 
is  formally the same upon replacing $\alpha'$ by $1/c_p$.
Thus, we obtain the eigenvalues:
\bea
 E^2 &=& p_y^2+ c_p \left( 2(2k+1)\ \big(1-\gamma\big)+\delta_p \right)\ 
\nonumber\\
&=& p_y^2-2bp_y(2k+1)+ c_p \left( 4k+2+ \delta_p \right)\ ,\qquad \gamma=\frac{bp_y}{c_p}\ .
\eea
To compute $\delta_p$, we follow the same procedure as in the string case.
Supersymmetry in the zero field limit requires, as before, $\delta_p\big|_{b=0}=-2$.
For finite $b$, we compute $\delta_p$ at for the ground state, $k=0$.
This corresponds to a rigid D$p$ brane, which can be dimensionally  reduced to a D-string along the directions $y_j$, $j=2,...,p$. 
The $k=0$ state then describes the state $S_{R}=1$, $S_{L}=-1$ having energy (\ref{dilaga}) (for this state, the energy does not depend on the tension; it corresponds to a point-like limit described by the effective field theory). 
This leads to $\delta_p=-2$ once again. Thus we have the general formula for quantum D$p$ brane 
states in this sector:
\be
 E^2 = p_y^2+ c_p \left( 2(2k+1)\ \big(1-\gamma\big) -2\right)\ ,
\ee
i.e.
\be\label{mako}
 E^2 = p_y^2+ \frac{R^{p-1} w}{l_s^{p+1} g_s} \left( 2(2k+1)\ \big(1-\gamma\big) -2\right) \ .
\ee
{}At large $k$, the energy (\ref{mako}) agrees with the energy formula  (\ref{spaEp}) of the classical pulsating D$p$ brane.

We see that a D$p$ brane with quantum numbers $n,\ k,\ w$ becomes tachyonic above a critical field
\be\label{fincri}
 R\, b_{\rm cr} = \frac{n}{4k+2} +\frac{2k}{2k+1} \frac{R^{p+1}w}{l_s^{p+1} g_s n}\ .
\ee
Thus, as  expected, a macroscopic D$p$ brane  can become tachyonic when magnetic fluxes are turned on.

\section{Discussion}

In this paper we have studied  instabilities of the Nielsen-Olesen type in backgrounds with
RR magnetic fluxes.
In our examples, gravitational back reaction to  the magnetic flux is incorporated exactly.

In compactifications having fluxes on cycles, fluxes are quantized in  units of the inverse volume of the cycle and
an interesting question is when to expect instabilities.
The mere existence of a single tachyon in the quantum spectrum of a theory implies that the vacuum is unstable. 
 It is obvious that, to linear order in the magnetic flux, the energy formula for a D$p$ brane will contain the standard gyromagnetic interaction.
Let us see this explicitly in a simple example.
Let $x^1,...,x^{p+2}$ be the coordinates of a rectangular torus, $T^{p+2}$.  We consider the constant field configuration
\be\label{artu}
A_{p+1}=f\ x_{1} \ dx_2\wedge ...\wedge dx_{p+2}\ .
\ee
Moreover, we choose a static gauge for the world-volume coordinates where $x_j= w_j R_j\sigma^j$, $j=3,...,p+2$, and let us first assume that $x_1,\ x_2$ only depend on $\tau $. Then the coupling to the gauge field in the D$p$ brane action becomes
\be\label{poty}
i\tau_p (2\pi)^p R_3...R_{p+2}\ w f\int d\tau x_1\p_\tau x_2 
\equiv i w B  \int d\tau x_1\p_\tau x_2  
\ee 
where $w=w_3...w_{p+2}$ . The flux quantization condition can be obtained by demanding that the generalized Wilson loop
\be
e^{i\tau_p\int A_{p+1}}\ ,
\ee
is single-valued under $x_1\to x_1+2\pi R_1$ (with periodic $\tau $ for a closed contour).
This implies
\be\label{fquan}
f=\frac{2\pi  k}{\tau_p {\rm Vol}(T^{p+2})}\ ,
\ee
where $k$ is an integer. This reproduces the result of \cite{Bousso:2000xa}.
Thus
\be
\label{Bquan}
B=\frac{2\pi k}{{\rm Vol}(T^2)}\ ,\qquad {\rm Vol}(T^2)=4\pi^2 R_1R_2\ .
\ee
The interaction (\ref{poty}) exhibits  a gyromagnetic coupling to the angular momentum of the brane in the plane $(x_1,x_2)$. In particular, when $p=1$, this is the same coupling that
appears in open string theory (the open string mass spectrum for magnetic field in toroidal directions was computed in \cite{Bachas:1995ik}).
There is no rotational symmetry in the 12 plane, but as usual spin
is defined for the Lorentz group acting on tangent space.
Because we assumed that $x_1,\ x_2$ only depend on $\tau$, the brane has only
orbital angular momentum. We can relax this assumption and let $x_1,\ x_2$
depend on some $\sigma^i$, in which case the brane will also carry spin and this term will give
rise to the standard gyromagnetic interaction
$ wB \big(2\ell +1 -2 S \big)$.
To linear order in the field $f$, for a brane in a quantum state of spin $S$, we would expect the mass formula to contain the terms
\bea
M^2 &=& M_0^2+ \sum_i \frac{n_i^2}{R_i^2} + \left(w \tau_p {\rm Vol}(T^{p})\right)^2-2w \tau_p {\rm Vol}(T^{p}) f J_{12}+...
\nonumber\\
&=&
M_0^2+ \sum_i \frac{n_i^2}{R_i^2} + \left(w \tau_p {\rm Vol}(T^{p})\right)^2+ (2\ell +1) B w-2 S B w+...
\label{gty}
\eea
where $\ell =0,1,2,...$ are the Landau levels and $M_0^2$ represents positive
contributions originating from oscillations and kinetic energy of the brane.
It must be noted that, for a D$p$ brane, there
are other linear couplings to the magnetic flux; even for the string, we have a coupling
to $\hat J_R-\hat J_L$.
Such coupling does not show up in this simple example
because of the particular ansatz we took for the brane embedding
(the coupling to $\hat J_R-\hat J_L$,  related to $x_1\partial_{\sigma_1} x_2-x_2\partial_{\sigma_1} x_1$, appeared  in the pulsating D brane discussed in section 3). 
The most general linear coupling to the flux is essentially read from (\ref{artu}) once the brane embedding is given.

However, this  example gives an easy insight into some important features: 1) It exhibits  the explicit dependence on the quantized flux in the gyromagnetic interaction. 2) In the exact quantum string spectrum (\ref{espiero}), the gyromagnetic coupling involving the winding charge appears upon expanding a positive definite term.
This exposes the fact that  $O(B^2)$ terms are important in order to make a statement about the positivity of $M^2$. Determining such $O(B^2)$ terms requires taking into account the gravitational back reaction.
This is a difficult problem in more general contexts, but the important point is that the linearized approximation already
identifies the precise  D brane states that can become light at critical fields.
3) Naively one might think that for sufficiently high spin, 
 the term $2 S Bw$ would render $M^2$ negative. But, of course,
the angular momentum gives,  in addition, a positive contribution to the kinetic energy, represented by $M_0^2$. In the string model, the oscillation/kinetic energy is represented by $2\hat N_L+2\hat N_R$,
and the highest spin for a given kinetic energy  is determined by the bound $|S_{L,R}|\leq \hat N_{L,R}+1$.

In our background magnetic flux, we have identified a tachyonic D brane state. It has zero angular momentum, but the linear coupling to the flux comes through $\hat J_R-\hat J_L$ (Right and Left being associated with $\big(\partial_\tau\pm \partial_{\sigma_1}\big) \varphi$).
At large quantum numbers, it admits a semiclassical description in terms of a pulsating 
D brane.
In general terms, even for supersymmetric configurations,  where
the spectrum is tachyon free and the vacuum is stable, we have seen that
certain high angular momentum states become very light at critical fields. 
A naive field theory analysis would ignore such D brane quantum states, because  they typically have huge energies of order $\tau_p {\rm Vol}(T_p) \omega $. However, we now see that at
a certain critical field, such states can be very light, with energies $E\ll 1/l_s$.
Near the critical point,  an effective field theory analysis should take into account these modes in order
to properly describe the low energy dynamics  of the system.\footnote{A recent discussion on brane effective field theory in background fluxes can be found in \cite{Polchinski:2015bea}.}
More generally, for any given flux configuration, one can easily compute the classical energy formula to linear order  in the fluxes  to see if the spectrum contains  a D brane state
with high quantum numbers that became light by the interaction with the flux.
In configurations with no residual supersymmetry, the appearance of D$p$ brane tachyonic modes above some critical fields seems to be a common feature, by the same mechanism  that induces  Nielsen-Olesen instabilities in quantum field theory.

Non-supersymmetric flux compactifications 
have been extensively used for the construction of inflationary models and  semi-realistic string compactifications \cite{Kachru:2003aw}. In these models, fluxes generate a superpotential which freezes all Calabi-Yau moduli and supersymmetry is typically broken by
the presence of anti-branes. An important question is whether some of these models could be affected by
tachyonic instabilities of the kind studied in this paper. 
To address this problem, a direct approach is to use the D$p$ brane Lagrangian (\ref{Infeld}) in the given background flux  and to compute the classical energy formula --analogue to (\ref{gty})-- for a D$p$ brane motion that maximizes the gyromagnetic interaction with low cost in oscillation/kinetic energy. Typically, these are rigidly rotating D$p$ branes  or pulsating Dp branes with no other oscillation modes turned on. A quick estimate of the expected order of the critical field can be obtained from (\ref{gty}), giving
$f\sim \frac{w}{2J} \tau_p {\rm Vol}(T^{p})$.\footnote{As discussed below (\ref{gty}), $J$ does not need to be the angular momentum.} This is consistent with
our examples, the BPS state  
 with energy (\ref{otroB}) or the pulsating $Dp$ brane with critical field (\ref{fincri}).
At stronger fields,  potential tachyonic instabilities can arise and the vacuum becomes unstable. This implies a bound for the maximum possible value for the quantized flux (\ref{fquan}),  roughly
$k \lesssim  \frac{c\tau_p^2}{4\pi} {\rm Vol}(T^{p+2}){\rm Vol}(T^{p})$, where $c$ is  a rational number  (for our examples, $w/J$ or $w/n$), which in a first approximation may be set to 1.\footnote{In particular, for an $H_3$ flux, this would give $k \lesssim (R/l_s)^4$.}
It would be extremely interesting to revisit the counting of 
 metastable string vacua with semi-realistic phenomenology (see e.g. \cite{Douglas:2003um}), in view of these new  constraints on the possible choices of flux.

\subsection*{Acknowledgements}
 
We would like to thank Arkady Tseytlin  and Diego Rodriguez-Gomez for useful comments.
We acknowledge financial support from projects  FPA2013-46570,  
 2014-SGR-104 and  MDM-2014-0369 of ICCUB (Unidad de Excelencia `María de Maeztu').



\begin{thebibliography}{20}

\bibitem{Nielsen:1978rm} 
  N.~K.~Nielsen and P.~Olesen,
  ``An Unstable Yang-Mills Field Mode,''
  Nucl.\ Phys.\ B {\bf 144}, 376 (1978).

\bibitem{Ambjorn:1988tm} 
  J.~Ambjorn and P.~Olesen,
  ``On Electroweak Magnetism,''
  Nucl.\ Phys.\ B {\bf 315}, 606 (1989).

\bibitem{Ambjorn:1988gb} 
  J.~Ambjorn and P.~Olesen,
  ``A Magnetic Condensate Solution of the Classical Electroweak Theory,''
  Phys.\ Lett.\ B {\bf 218}, 67 (1989)
  Erratum: [Phys.\ Lett.\ B {\bf 220}, 659 (1989)].

\bibitem{Ambjorn:1989bd} 
  J.~Ambjorn and P.~Olesen,
  ``A Condensate Solution of the Electroweak Theory Which Interpolates Between the Broken and the Symmetric Phase,''
  Nucl.\ Phys.\ B {\bf 330}, 193 (1990).

\bibitem{Ferrara:1993sq} 
  S.~Ferrara and M.~Porrati,
  Mod.\ Phys.\ Lett.\ A {\bf 8}, 2497 (1993)
  [hep-th/9306048].

\bibitem{Bachas:1995ik} 
  C.~Bachas,
  ``A Way to break supersymmetry,''
  hep-th/9503030.

\bibitem{Russo:2005za} 
  J.~G.~Russo, ``Strong magnetic limit of string theory,''
  JHEP {\bf 0506}, 005 (2005)
  [hep-th/0504187].


\bibitem{RTexactly} 
  J.~G.~Russo and A.~A.~Tseytlin,
  ``Exactly solvable string models of curved space-time backgrounds,''
  Nucl.\ Phys.\ B {\bf 449}, 91 (1995)
  [hep-th/9502038].

\bibitem{RTflux} 
 J.~G.~Russo and A.~A.~Tseytlin,
  ``Magnetic flux tube models in superstring theory,''
  Nucl.\ Phys.\ B {\bf 461}, 131 (1996)
  [hep-th/9508068].
  
\bibitem{Kiritsis:1995iu} 
  E.~Kiritsis and C.~Kounnas,
  ``Infrared behavior of closed superstrings in strong magnetic and gravitational fields,''
  Nucl.\ Phys.\ B {\bf 456}, 699 (1995)
  [hep-th/9508078].

\bibitem{RTmagnetic}
 J.~G.~Russo and A.~A.~Tseytlin,
  ``Magnetic backgrounds and tachyonic instabilities in closed superstring theory and M theory,''
  Nucl.\ Phys.\ B {\bf 611}, 93 (2001)
  [hep-th/0104238].
  
\bibitem{Minwalla:2003hj} 
  S.~Minwalla and T.~Takayanagi,
  ``Evolution of D branes under closed string tachyon condensation,''
  JHEP {\bf 0309}, 011 (2003)
  [hep-th/0307248].

\bibitem{Diengo}
  R.~Iengo, J.~Lopez Carballo and J.~G.~Russo,
  ``Strings and D-branes in a supersymmetric magnetic flux background,''
  JHEP {\bf 0708}, 047 (2007)
  [arXiv:0707.0455 [hep-th]].

\bibitem{RTfluxbranes}
 J.~G.~Russo and A.~A.~Tseytlin,
  ``Supersymmetric fluxbrane intersections and closed string tachyons,''
  JHEP {\bf 0111}, 065 (2001)
  [hep-th/0110107].


\bibitem{Bousso:2000xa} 
  R.~Bousso and J.~Polchinski,
  ``Quantization of four form fluxes and dynamical neutralization of the cosmological constant,''
  JHEP {\bf 0006}, 006 (2000)
  [hep-th/0004134].

  
\bibitem{Polchinski:2015bea} 
  J.~Polchinski,
  ``Brane/antibrane dynamics and KKLT stability,''
  arXiv:1509.05710 [hep-th].


\bibitem{Kachru:2003aw} 
  S.~Kachru, R.~Kallosh, A.~D.~Linde and S.~P.~Trivedi,
  ``De Sitter vacua in string theory,''
  Phys.\ Rev.\ D {\bf 68}, 046005 (2003)
  [hep-th/0301240].

\bibitem{Douglas:2003um} 
  M.~R.~Douglas,
  ``The Statistics of string / M theory vacua,''
  JHEP {\bf 0305}, 046 (2003)
  [hep-th/0303194].

\end{thebibliography}
\end{document}